\documentclass[%
 reprint, amssymb, amsmath, aip, sd, 
]{revtex4-1}

 \usepackage{SIunits}
 \usepackage{CJK}

 \usepackage{upgreek}
\usepackage{graphicx}
\usepackage{epstopdf}

\usepackage{float}
\usepackage{color}
\usepackage{docs}%
\usepackage{bm}%
\usepackage[colorlinks=true,linkcolor=blue,citecolor=blue]{hyperref}%
\usepackage{color, soul, framed}

\usepackage{lineno}

\expandafter\ifx\csname package@font\endcsname\relax\else
 \expandafter\expandafter
 \expandafter\usepackage
 \expandafter\expandafter
 \expandafter{\csname package@font\endcsname}%
\fi
\begin{document}

\title{Varied fusion reaction probability induced by ion stopping modification in laser-driven plasma with different temperature}%

\author{Yihang Zhang}%
\affiliation{Beijing National Laboratory for Condensed Matter Physics, Institute of Physics, Chinese Academy of Sciences, Beijing 100190, China}%
\affiliation{School of Physical Sciences, University of Chinese Academy of Sciences, Beijing 100049, China}%

\author{Zhe Zhang}%
 \email{zzhang@iphy.ac.cn}
\affiliation{Beijing National Laboratory for Condensed Matter Physics, Institute of Physics, Chinese Academy of Sciences, Beijing 100190, China}%

\author{Baojun Zhu}%
\affiliation{Beijing National Laboratory for Condensed Matter Physics, Institute of Physics, Chinese Academy of Sciences, Beijing 100190, China}%
\affiliation{School of Physical Sciences, University of Chinese Academy of Sciences, Beijing 100049, China}%

\author{Weiman Jiang}%
\affiliation{Beijing National Laboratory for Condensed Matter Physics, Institute of Physics, Chinese Academy of Sciences, Beijing 100190, China}%
\affiliation{School of Physical Sciences, University of Chinese Academy of Sciences, Beijing 100049, China}%

\author{Lei Cheng}%
\affiliation{Beijing National Laboratory for Condensed Matter Physics, Institute of Physics, Chinese Academy of Sciences, Beijing 100190, China}%
\affiliation{School of Physical Sciences, University of Chinese Academy of Sciences, Beijing 100049, China}%

\author{Lei Zhao}%
\affiliation{Department of physics, College of Science, China University of Mining and Technology, Beijing 100083, China}%

\author{Xiaopeng Zhang}%
\affiliation{Key Laboratory for Laser Plasmas (Ministry of Education) and School of Physics and Astronomy, Shanghai Jiao Tong University, Shanghai 200240, China}%
\author{Xu Zhao}%
\affiliation{Key Laboratory for Laser Plasmas (Ministry of Education) and School of Physics and Astronomy, Shanghai Jiao Tong University, Shanghai 200240, China}%
\affiliation{Collaborative Innovation Centre of IFSA (CICIFSA), Shanghai Jiao Tong University, Shanghai 200240, China}%
\author{Xiaohui Yuan}%
\affiliation{Key Laboratory for Laser Plasmas (Ministry of Education) and School of Physics and Astronomy, Shanghai Jiao Tong University, Shanghai 200240, China}%
\affiliation{Collaborative Innovation Centre of IFSA (CICIFSA), Shanghai Jiao Tong University, Shanghai 200240, China}%

\author{Bowei Tong}%
\affiliation{Department of Astronomy, Beijing Normal University, Beijing 100875, China}%
\author{Jiayong Zhong}%
\affiliation{Department of Astronomy, Beijing Normal University, Beijing 100875, China}%
\affiliation{Collaborative Innovation Centre of IFSA (CICIFSA), Shanghai Jiao Tong University, Shanghai 200240, China}%

\author{Shukai He}%
\author{Feng Lu}%
\author{Yuchi Wu}%
\author{Weimin Zhou}%
\author{Faqiang Zhang}%
\author{Kainan Zhou}%
\author{Na Xie}%
\author{Zheng Huang}%
\author{Yuqiu Gu}%
\affiliation{Science and Technology on Plasma Physics Laboratory, Research Center of Laser Fusion, CAEP, Mianyang, Sichuan 621900, China}%

\author{Suming Weng}%
\affiliation{Key Laboratory for Laser Plasmas (Ministry of Education) and School of Physics and Astronomy, Shanghai Jiao Tong University, Shanghai 200240, China}%
\affiliation{Collaborative Innovation Centre of IFSA (CICIFSA), Shanghai Jiao Tong University, Shanghai 200240, China}%

\author{Miaohua Xu}%
\author{Yingjun Li}%
\affiliation{Department of physics, College of Science, China University of Mining and Technology, Beijing 100083, China}%

\author{Yutong Li}%
 \email{ytli@iphy.ac.cn}
\affiliation{Beijing National Laboratory for Condensed Matter Physics, Institute of Physics, Chinese Academy of Sciences, Beijing 100190, China}%
\affiliation{School of Physical Sciences, University of Chinese Academy of Sciences, Beijing 100049, China}%
\affiliation{Songshan Lake Materials Laboratory, Dongguan, Guangdong 523808, China}%

\begin{abstract}
The dynamics of nuclear reaction in plasma is a fundamental issue in many high energy density researches, such as the astrophysical reactions and the inertial confinement fusion. 
The effective reaction cross-sections and ion stopping power in plasma need to be taken into account to analyze the reactivity. 
In this research, we have experimentally investigated the from D-D reactions (D + D $\rightarrow$ $^3$He +n) from interactions between deuteron beams and deuterated polystyrene (CD) plasma, driven by two laser pulses respectively. 
The neutron yields, plasma density and deuteron energy loss in plasma have been measured, and the plasma temperature and deuteron stopping power have been analyzed from simulations.
It is shown that, compared with a cold target, the reaction probability in plasma conditions can be enhanced or suppressed, which is ascribed to the deuteron stopping power modifications in plasma.
In hotter CD plasma, the energy loss of moderate energetic deuterons reduces, which leads to higher D-D reaction probability,
while the contrary happens in colder plasma.
This work provides new understanding of fusion reactions in plasma environment.
\end{abstract}
\maketitle

The rapid development of the high-power laser technologies enabled research on
nuclear reactions in laser-plasma interactions, which has aroused comprehensive concerning and investigation \cite{voronchev2000nuclear,mckenna2003demonstration,ditmire1999nuclear,ledingham2000photonuclear}.
From neutronic reactions driven by high power lasers \cite{ditmire1999nuclear,roth2013bright,pomerantz2014ultrashort,kar2016beamed,abe2017production}, bright neutron sources have various promising applications in high resolution radiography \cite{loveman1995time,Anderson2009Neutron} and nondestructive treatment \cite{barth2012current}.
Thermonuclear reactions in inertial confinement fusion (ICF) \cite{nuckolls1972laser,Lindl1995Development} initiated by lasers can be expected as a possible attractive energy source of future environment.
Besides particle and energy generation, physical research on nuclear astrophysical reactions from laser-plasma interactions is a suitable approach to explore primordial nucleosynthesis in the universe \cite{CyburtBig,RyanPrimordial,hou2017non}.
Since most of the reactions take places in plasma, for laser-driven fusions, research on dynamics of plasma nuclear reaction is essential and strongly required.

Previous experiment \cite{LabauneFusion} has shown the proton-boron reaction yields in plasma can be improved by orders of magnitude compared to the one in a solid target.
However, the mechanisms of the increased fusion probability have not been well-understood, though alluded to abatement of proton stopping power in ionized matter.
For ion-induced fusion reactions, correction of ion stopping power in plasma affects the collision energy and reheating process of ions, then the reaction rate is changed.
Particularly, ion stopping power plays an important role in the ignition threshold analysis and the design of ICF \cite{Hora2008}.
In general, ion stopping in plasma is an comprehensive effect of small-angle collisions, large-angle scattering, as well as quantum degeneracy\cite{LiCharged,Brown2005Charged,CayzacPredictions}, with fundamental parameters as plasma density, temperature, ionization states, etc.
It is necessary to establish accurate models of ion stopping in plasma.
Additionally, varying electrical and magnetic fields and filamentation instability may be generated through Ohmic heating processes in the presence of an intense ion beam, which affects ion transportation and energy deposition \cite{askar1994magnetic,kim2015self} in turn.
Those processes and detailed mechanisms caused by plasma peculiarities can modify fusion probability, and need to be accurately analyzed in nuclear reaction dynamics research.

To investigate the nuclear reactions in plasma, simultaneous diagnoses for reaction yields, ion energy loss and plasma parameters are necessary.
In this work, we measured the three metrics in the interactions of an energetic deuteron beam and deuterated plasma driven by two separated laser pulses.
Different from Ref. 14, our results show the reaction probability variation is not always positive from a cold target to plasma.
Decrease (or increase) of the deuteron stopping power can lead to larger (or smaller) reaction probability in plasma with different temperature.

\begin{figure*}[!hbt]
\centering
\includegraphics[width=12cm]{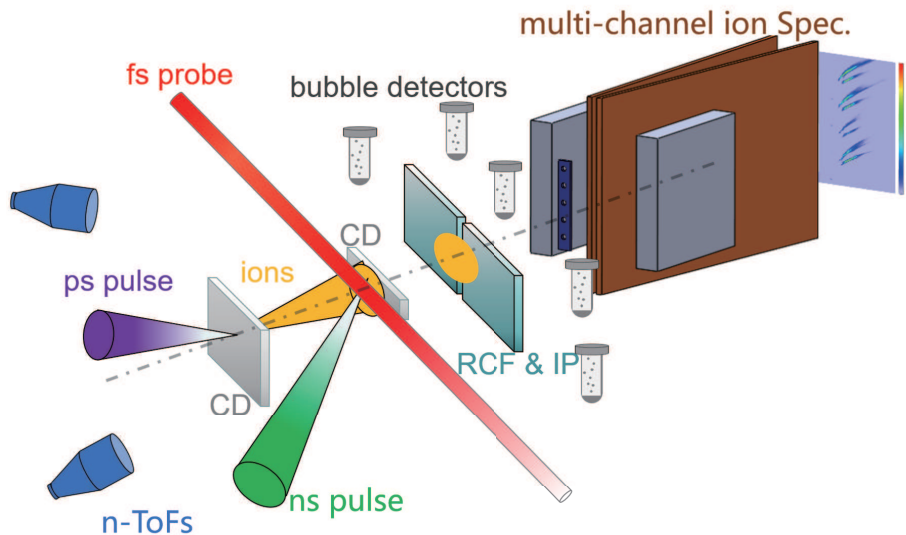}
\caption{\label{fig1}The schematic diagram of the experimental setup. 
Two separated CD foils were irradiated by ns and ps pulses respectively, and D-D reactions took places in the interactions between the deuteron beam and CD plasma.
Bubble and nToF detectors were set at different angles in the horizontal plane to measure the neutron yields.
A fs pulse was used as a probe for plasma density diagnosing.
The spatial beam profile of ions was measured by the RCF stack. 
Through the stack slit, the ions entered the ARTPS through multi-pin-hole channels and their tracks were recorded by the image plate (see the coloured map).
}
\end{figure*}

The experiment was carried out at the XG\textrm{-}III laser facility \cite{zhu2017xingguang}.
Figure~\ref{fig1} is a schematic diagram of the experimental geometry.
Nanosecond and picosecond laser pulses were employed to irradiate at two parallel deuterated polyethylene (CD) foils with a separation of 4 mm, respectively. 
The thickness of CD targets was 10 $\upmu$m.
A nanosecond (ns) pulse was used as a pre-heater to generated a CD plasma.
The pulse energy was 100 J and the intensity was $1\times 10^{15}$ W cm$^{-2}$, in a duration of 1 ns and a focal spot of 150 $\upmu$m full width at half-maximum (FWHM) diameter, 
Afterword, a picosecond (ps) pulse irradiated at the other CD target and proton and deuteron beams emitted through TNSA mechanism \cite{wilks2001energetic,zepf2001fast}.
The ps pulse had a wavelength of 1.05 $\upmu$m and the incident angle was $13^{\circ}$ relative to target normal in the horizontal plane.
The energy of the pulse was 100 J and the intensity was  $1.4 \times 10^{19}$ W cm$^{-2}$, in a duration of 0.9 ps and a focal spot of 20 $\upmu$m FWHM diameter.
The time delay between the two pulses was in 2 ns with a jitter less than 100 ps. 
There was a 1 $\upmu$m Al foil (not shown in Fig.~\ref{fig1} for brevity) in between the two targets to block the plasma expansion and protect the sheath field for ion acceleration.
When the deuteron beam interacted with the CD plasma, D-D reactions (D + D $\rightarrow$ $^3$He + n) were induced, as happened in the pitcher-catcher scheme \cite{roth2013bright,maksimchuk2013dominant}.  
Different plasma conditions which the deuteron beam experienced could be controlled by the adjusting the pre-heating time delay.

The reaction yields, plasma densities and deuteron energy loss were measured simultaneously.
A couple of BD-PND bubble detectors \cite{LewisReview,Zhao2018Laser} and neutron Time-of-Flight (nToF) \cite{GlebovThe,GlebovTesting,Hatarik2014A} detectors were employed to diagnose D-D neutron yields and angular distributions.
The plasma density at the target front surface irradiated by the ns pulse was diagnosed by phase imaging of a femtosecond (fs) pulse with 2 J in 40 fs transmitted from the plasma.
The time delay of the fs pulse after the ns one can be adjusted within 2.5 ns, with a jitter better than 100 ps.
The two-dimensional profile of the ion beam transmitted from the catcher target was measured by MD-V3 Radiochromic Film (RCF) \cite{vatnitsky1997radiochromic,snavely2000intense} stacks. 
The stacks had a 1-mm-wide slit making an angular-resolved Thomson parabola spectrometer (ARTPS) \cite{zhang_angular-resolved_2018} exposed to measure ion angular spectra at the same shot. 
Dispersed traces of protons and deuterons were recorded by a BAS-TR image plate (IP) \cite{alejo2014characterisation} at the end of the ARTPS.

\begin{figure*}[!hbt]
\centering
\includegraphics[width=15cm]{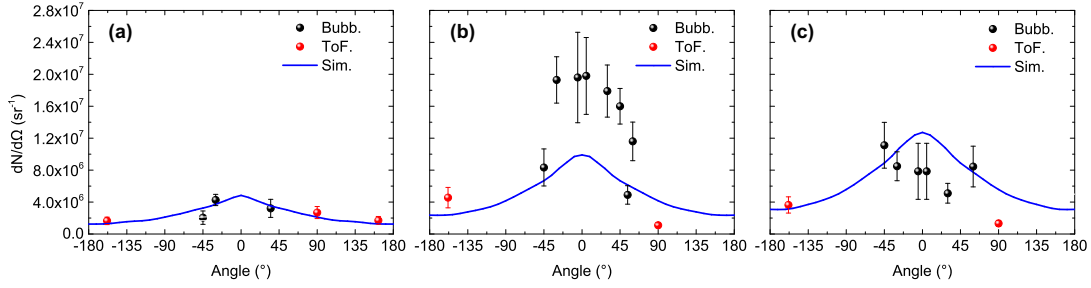}
\caption{\label{fig2}Neutron angular distribution without (a) and with 0.2 ns (b) and 1.8 ns (c) pre-heating. 
The black and red solid circles are from measurements of the bubble and ToF detectors, respectively.
Simulated neutron angular distributions are shown as the blue curves.
The simulation produces a reference which well predicts the results for the cold target as shown in panel (a).
For the short- and long-delay pre-heating illustrated in panels (b) and (c), there are increase and decline of reaction yields, respectively, compared with a cold target. 
}
\end{figure*}

To present a comparison of reaction yields from plasma and a cold target, Monte-Carlo (MC) simulations containing data of ion stopping power and reaction cross section in cold targets has been employed.
Importing the measured angular and energy distribution of deuteron beam on each shot, MC simulations set a reference for D-D neutron production in a cold CD target.
Figures~\ref{fig2} (a) plots the neutron angular distribution generated from a cold target without pre-heating.
0 degree is the direction of the ion propagation (target normal).
The black and red solid circles represent results from bubble and nToF detectors, respectively, with error bars summing systematic and statistical errors in quadrature.
Shown as the blue curve in Fig.~\ref{fig2} (a), the simulation well predicts both of the angular distribution and neutron yield for the cold target.
However, with 0.2 ns pre-heating (from the rising edge of the ns pulse to the ps one irradiating at the targets) there is an enhancement of neutron yields especially for forward neutrons, while with 1.8 ns pre-heating the neutron yields reduces, compared to those in the cold target illustrated by the blue curves [see Fig.~\ref{fig2} (b) and (c)].
Different pre-heating time-delay leads to different plasma environment which deuteron beams experienced.
The results of the reaction yields indicate that the plasma condition makes the reaction probability per deuteron modified.

\begin{figure}[!hbt]
\centering
\includegraphics[width=8cm]{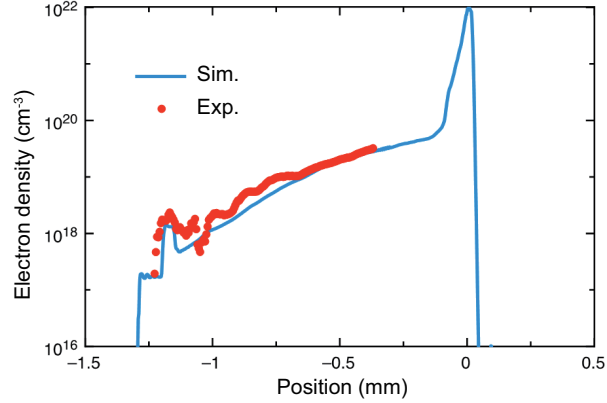}
\caption{\label{fig3} Comparison between measured (red circles) and simulated (blue curve) free electron density profiles along the axis of ion propagation (target normal).
}
\end{figure}

In order to further clarify the physical processes associated with the modified reaction probability presented above, we have investigated the plasma parameters.
A 400 nm probe beam with an energy of 1.5 mJ and duration of 40 fs has propagated parallelly to the CD foil surface, to measure the plasma density at the front side of the catcher target.
A 2D phase map of the probe beam through the plasma was recorded by the SID4-HR sensor \cite{SID4-HR}, and the plasma density was inverted.
Radiation-hydrodynamics simulation by the Flash code \cite{FryxellFLASH,DubeyIMPOSING} has been used to simulate the CD plasma evolution driven by the ns pulse.
Typical longitudinal distribution of plasma density at 2 ns after the rising edge of the ns pulse in Fig.~\ref{fig3} shows nice agreement between the measurement and simulation in the detectable density range.
Besides, the FLASH simulation also provides temperature and density for both electrons and ions, as well as ionization states of the plasma. 
The catcher target is consist of three parts along the deuteron beam path, fully-ionized weakly-coupled plasma with low density, partly-ionized moderately-coupled plasma with high density and unionized matter with solid density.
The fusion probability is a path-length-integrated effect of deuteron beam transportation in the catcher target.

\begin{figure}[!hbt]
\centering
\includegraphics[width=9cm]{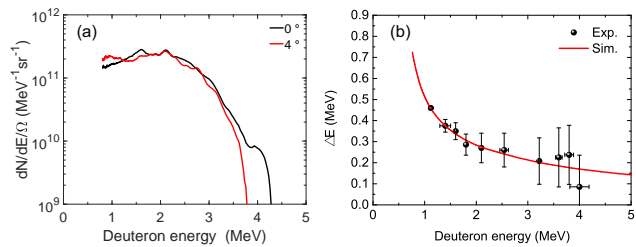}
\caption{\label{fig4}
(a) Deuteron energy spectra in 0 $^\circ$ (black curve) and 4 $^\circ$ (red curve) from a pitcher CD target driven by the ps laser pulse.
(b) Deuteron energy loss in the cold catcher target. 
The data points in back circles are deduced from measurements of the ARTPS, consistent with the results from MC simulations shown as the red curve. 
The error bars in the vertical axis are given by the energy resolution of the ARTPS.
The error bars in the horizontal axis are owing to the energy deviation determined by the same deuteron numbers between the spectra in panel (a).
}
\end{figure}

The deuteron energy loss in the catcher target has been measured by the ARTPS.
The diameter of the entrance pin-holes is 60 $\upmu$m to achieve an energy resolution of $\delta E/E=0.02$ for 1 MeV deuterons. 
The spectrometer has a detective deuteron energy threshold of 0.7 MeV.
Firstly, to characterize the accelerated deuteron beam, test shots have been taken with only the pitcher target and ps pulse.
Deuterons entered the ARTPS through 4 pin-holes in a vertical row, at angles from -2$^\circ$ to 4$^\circ$, with respect to the target normal.
From the parabolic traces recorded on the IP, angular-resolved deuteron energy spectra can be resolved\cite{zhang_angular-resolved_2018}.
As seen in Fig. ~\ref{fig4} (a), for the majority of the energy range ($\textless$ 3.5 MeV), quasi-identical deuteron spectra are acquired at 0 and 4 degrees, except for a small range around the cutoff energy.
With the catcher target introduced, without pre-heating, energy spectra of deuterons after the cold target can be determined.
The width of the catcher target in the vertical direction is 0.5 mm, and the distance between the pitcher and catcher is 4 mm.
According to the geometry, deuterons measured at 4$^\circ$ have no block or energy loss after emitting from the pitcher.
Considering the small differences on spectra,
deuterons with energy less than 3.5 MeV from 4$^\circ$ and 0$^\circ$ can be regarded as beams incident at and transmitted from the catcher target.
In the 10-$\upmu$m-thick catcher, the mean lateral straggling of a 1.2 MeV deuteron is 0.08 $\upmu$m, and the corresponding displacement at the ARTPS entrance is 60 $\upmu$m within the pin-hole diameter.
So the number difference of the incident and transmitted beams can be neglected for deuterons with incident energy above 1.2 MeV.
In this energy range (1.2 $\sim$ 3.5 MeV), deuteron energy loss can be estimated by the energy difference at same deuteron number between the two spectra from 4$^\circ$ and 0$^\circ$.
Fig. ~\ref{fig4} (b) illustrates the deuteron energy loss in a cold CD target, comparing results derived from the ARTPS measurement and MC simulations.
Taking the error bars into account, the measured energy loss of deuterons in the cold target agrees well with the simulated results.

\begin{figure}[!hbt]
\centering
\includegraphics[width=9cm]{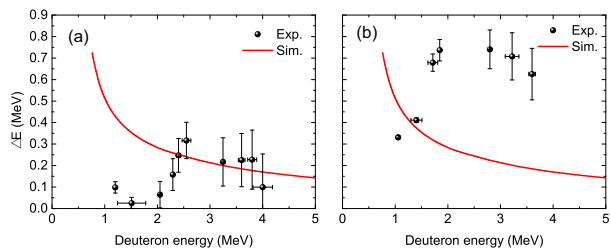}
\caption{\label{fig5} 
Deuteron energy loss in 0.2 ns (a) and 1.8 ns (b) pre-heated plasma as functions of incident energy.
The black circles show the energy loss from ARTPS measurements, while the red curve shows the values from MC simulations for the cold target. 
The error bars have the same sources with Fig.~\ref{fig4} (b).
Compared to the cold target, 
for 0.2 ns pre-heating the energy loss for deuterons with energy less than 2.4 MeV gets smaller,
while for 1.8 ns pre-heating the energy loss becomes stronger for deuterons with energy above 1.5 MeV.
}
\end{figure}

After checking and assuring the results from the cold target,
we introduced the ARTPS to diagnose deuteron energy loss in plasma.
For 0.2 ns and 1.8 ns pre-heating, the deuteron energy loss from the ARTPS  is shown as the black circles in Fig.~\ref{fig5} (a) and (b), respectively.
The red curve references energy loss in the cold CD target from the MC simulations.
For the shorter time delay, the energy loss of deuterons in 2.5 $\sim$ 4 MeV has fairly little difference with the one in the cold target, but
for less-energetic part the energy loss is smaller, as shown in Fig.~\ref{fig5} (a).
For the longer time delay the energy loss of high-energy deuterons is larger than the prediction of the cold target, as shown in Fig.~\ref{fig5} (b).

\begin{figure}[!hbt]
\centering
\includegraphics[width=8cm]{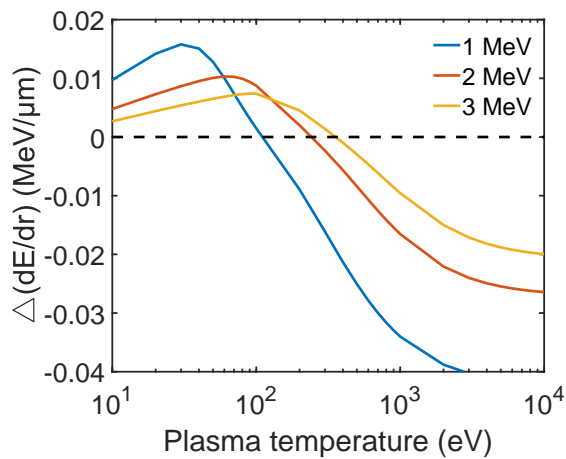}
\caption{\label{fig6} 
Difference between deuteron stopping power in CD plasma and the one in a cold CD target ($\Delta\frac{\mathrm{d}E}{\mathrm{d}r}=(\frac{\mathrm{d}E}{\mathrm{d}r})_{plasma}-(\frac{\mathrm{d}E}{\mathrm{d}r})_{solid}$), as functions of plasma temperature for deuterons with energy of 1, 2 and 3 MeV.
The stopping power in plasma is calculated from the BPS formalism, while one in the cold target from MC simulations.
The CD plasma and cold CD have the same density of 1.1 g/cm$^3$.
}
\end{figure}

The deuteron arriving time at the catcher target varies under different pre-heating delay.
With 0.2 ns pre-heating, for deuterons with energy from 0.1 to 10 MeV, which has cross sections of D-D reactions above 0.02 barn, the arriving time is from 0.3 ns to 1.3 ns since the rising edge of the ns pulse.
This period overlaps the pre-heating pulse from 0 to 1 ns, and the plasma can be heated and reach high temperature.
However, the arriving time is from 1.9 ns to 2.9 ns for those deuterons under 1.8 ns pre-heating, when the heating has already finished, so the plasma has been cooled down to some extent.
The deuteron stopping power is highly sensitive to plasma temperature variation \cite{Mehlhorn1981finite,li1993charged}.
According to the Coulomb interaction theory and the Fokker-Planck formulation, the maximum stopping power of an ion beam occurs when its projectile velocity $v_b$ is close to the thermal velocity of the free plasma electrons $v_{th}$.
Among the ion-stopping models, the Brown-Preston-Singleton (BPS) formalism \cite{brown_charged_2005} has been validated experimentally in a relatively wide range of $v_b/v_{th}$ ratio \cite{frenje2019experimental}.
Deuteron stopping power difference $\Delta\frac{\mathrm{d}E}{\mathrm{d}r}$ between the CD plasma and cold CD varying with background temperature are calculated using  the BPS formalism, which is shown in Fig.~\ref{fig6}.
The figure shows for a deuteron beam with a specific kinetic energy, there is an plasma temperature threshold above which the stopping power becomes smaller than that in a cold target.
The threshold gets larger for higher energy deuterons.
Under 0.2 ns pre-heating, when deuterons with energy above 2.4 MeV arrives at the catcher target, the latter has not been entirely ionized or efficiently heated, so the deuteron stopping power in this energy range hardly shows differences compared with the cold target.
When deuterons with energy less than 2.4 MeV arrive, the plasma has been heated up to 300 eV and grows over time, so the stopping power reduces [see Fig.~\ref{fig5} (a)].
For 1.8 ns pre-heating the plasma is lower than 150 eV when deuterons arrive.
As a result, for the deuterons above 1.5 MeV, the stopping power is higher than the prediction of the cold target [see Fig.~\ref{fig5} (b)].

Along an infinitesimal propagation distance d$r$, the reaction probability (reaction numbers generated with per unit number of incident ions) can be written as,
\begin{eqnarray}
\label{eq:eq1}
    \frac{\mathrm{d}N_{fus}}{\mathrm{d}N_b}&=&n_{p}\sigma(E)\mathrm{d}r,
\end{eqnarray}
where $n_p$ is the number density of reacting nuclei in the plasma, and $\sigma$ is the reaction cross section related to collision energy $E$.
Normalized by the target density, the ion stopping power $\epsilon(E)$ = $\frac{\mathrm{d}E}{n_{p}\mathrm{d}r}$, then equation (\ref{eq:eq1}) can be written as
\begin{eqnarray}
\label{eq:eq2}
\frac{\mathrm{d}N_{fus}}{\mathrm{d}N_b}&=&\frac{ {\sigma(E)}}{{\epsilon(E)}}\mathrm{d}E.
\end{eqnarray}
The reaction probability is inversely proportional to the ion stopping power.
As a Maxwellian-energy-distributed deuteron source from TNSA, the number becomes lower against the energy exponentially.
For 0.2 ns pre-heating, 
the energy loss of a great number of less-energetic deuterons ($\textless$ 2.4 MeV) in the plasma reduces, and it leads to higher efficient D-D collision energy compared with the cold target. 
According to equation (\ref{eq:eq2}) there is a neutron yield enhancement, consistent with the result shown in Fig. \ref{fig2} (b).
On the other hand, for 1.8 ns pre-heating, the stopping power of high-energy deuterons is larger than that in the cold target [see Fig. \ref{fig5} (b)], so the inhibition of deuteron propagation suppresses energetic collisions.
There used to be high reactivity originating from the barrier penetrability for high-energy deuterons.
As a result, the decrease of reaction probability induced by those high-energy deuterons leads to the neutron yield reduction [see Fig. \ref{fig2} (c)]. 

There are other physical processes could possibly impact on the fusion reaction probability in plasma.
The reaction cross section in plasma differs from the one in the cold target due to the electron screening effect, which would also lead to reaction probability variation according to equation (\ref{eq:eq1}).
Nevertheless, 
for D-D reactions, this effect shows significance only at center-of-mass energy lower than 10 keV \cite{Barker2016Electron}.
So the correction of cross section is negligibly small for reactions in energetic beam-target interactions explored in this work rather than that in thermonuclear reactions.
Otherwise, as an intense beam current transports in a high-density matter, electro-magnetic fields and Ohmic heating effects can be excited \cite{Kim2015PRL, kim2016varying}.
However, that is not the case in the present experiment where the deuteron beam current on the catcher target is less than 10$^6$ A/cm$^2$ deduced from the ARTPS measurement.
Further, during the laser heating, noncollinearity between density and temperature gradients in the high-energy-density plasma could drive strong magnetic fields \cite{Nilson2006Magnetic}.
Ion beams tend to be dispersed and pinched under the fields.
However, regarding to the hydrodynamics simulation, for a 0.7 MeV deuteron penetrating the 10-$\upmu$m-thick target the maximum displacement is only 0.1 $\upmu$m , under a 17 T magnetic field in 20 $\upmu$m around the target front surface.
So the dispersion of deuteron beam in the plasma can be neglected.
Anyhow, these potential mechanisms would be inconsistent with the considerable modifications on reaction yields observed in Fig. \ref{fig2} (b) and (c).
In general, corrections of deuteron stopping power in plasma can explain the fusion probability variation well.
Actually, further enhancement and optimization of fusion reaction probability can be investigated through restriction of ion stopping power by adjusting plasma temperature.

In summary, to study the fusion reactions in plasma, D-D reaction yields together with the plasma parameters and deuteron energy loss have been investigated simultaneously.
Compared with a cold target, variation of reaction probability in plasma have been observed, which is mainly ascribed to the modification of the deuteron stopping power depending on the plasma temperature.
This approach suggests new understanding of fusion  reactions in plasma condition.
Additionally, the experimental scheme may also enable a cross-section diagnosis for fusion reactions in plasma, by measuring the ion stopping power, plasma parameters and reaction yield on a single shot.
In particular, the abatement of deuteron energy loss in hot plasma can lead to higher reaction probability, which provides an efficient method for neutron sources driven by lasers. 


\bibliography{ref}
\end{document}